\documentclass[pra,aps,preprint]{revtex4}
\usepackage{amsmath}
\usepackage{amssymb}
\usepackage{graphicx}

\begin{document}

\title{Stability of germanene under tensile strain}

\author{T. P. Kaloni and U. Schwingenschl\"ogl}

\email{udo.schwingenschlogl@kaust.edu.sa,+966(0)544700080}

\affiliation{Physical Science \& Engineering Division, KAUST, Thuwal 23955-6900, Kingdom of Saudi Arabia}

\begin{abstract}
The stability of germanene under biaxial tensile strain and electronic properties 
are studied by means of density functional theory based calculations. Our results show that up to
16\%  biaxial tensile strain germanene lattice is stable and the Dirac cone shifts towards higher 
energy range with respect to the Fermi level as a result $p$-doped Dirac states are achieved. The realization of 
the $p$-doped Dirac states are due to the weakening of the Ge$-$Ge bonds and reduction of hybridization 
with the strain. We therefore calculate the phonon spectrum to demonstrate that the germanene is stable 
up to 16\% under biaxial tensile strain. Our calculated Gr\"uneisen parameter shows the similar trend to 
silicene and different trend to graphene under small biaxial tensile strain.
\end{abstract}

\maketitle

Graphene is a two-dimensional (2D) honeycomb lattice of carbon atoms, currently a material 
of interest for many researcher due to the fact that its unique electronic properties, 
which is being proposed to be a great potential for future nanoelectronic applications \cite{geim}. 
The mass production and band gap opening are real challenges as a result searching of a new materials which can be 
a counterpart of graphene is highly demanded. Recent years, the electronic properties of 2D hexagonal 
silicon and germanium also named as silicene and germanene, respectively, have been proposed as a potential 
alternative of graphene \cite{Topsakal}. Experimentally, it has been demonstrated Ag and 
ZrB$_2$ substrates \cite{padova,vogt,ozaki} can be used to grow the silicene. However, the free standing silicene 
and germanene are not realized so far. C, Si, and Ge belongs to the same group in the periodic table whereas, Si and Ge have 
a larger ionic radius, which promotes $sp^3$ hybridization. The mixture of $sp^2$ and $sp^3$ hybridization in 
silicene and germanene results in a prominent buckling (0.46 \AA\ and 0.68 \AA\ for silicene and germanene) as 
compared to graphene, which opens an electrically tunable band gap \cite{falko,Ni}. As a consequences, its a huge advantage 
as compared to graphene. 

Germanene was proposed to be a poor metal \cite{Topsakal}. In this study, the authors ignored 
the intrinsic spin orbit coupling. The magnitude of the spin orbit coupling is significantly larger in germanene and can not be 
neglected as it is a materials property. It was also noted that in-plane biaxial compressive strain turns 
germanene into a gapless semiconductor, by remain intact the linear energy dispersions at the 
K and K$'$ points \cite{Houssa}. 
The magnitude of the intrinsic spin orbit coupling in Ge (6.3 meV) is stronger than that of Si (4 meV) and 
C (1.3 $\mu$eV) atoms \cite{Liu1}. It has been demonstrated that germanene can be a good candidate for 
the quantum spin hall effect with a sizable band gap at the Dirac points due to 
stronger spin orbit coupling and the higher buckling as compared to silicene \cite{Liu,Liu1}. 
As a result, germanene can be a potential candidate for constructing promising spintronic devices. 
The absorption of F, Cl, Br and I has been studied \cite{ma} and found that the intrinsic spin orbit 
coupling band gap in germanene is enhanced by absorption up to 162 meV, clearly higher 
than that for pristine germanene.

Strain takes play a role when a crystal is compressed (stretched) from the equilibrium. The strain can affect the device 
performance, it can be applied intentionally to improve mobility. The biaxial tensile strain modify the crystal phonons, 
which usually resulting in mode softening. The rate of these changes is determined by the Gr\"uneisen parameters, which 
also can determine the thermomechanical properties \cite{Mohiuddi}. 
Graphene preserves the zero gap semiconducting nature even under huge strain of about 30\% \cite{choi}. 
However in silicene, the lattice stable up to 17\% \cite{kaloni} and shows self hole doped Dirac states \cite{Liu2}. 
Hence, it is an important issue whether the stability and electronic properties are modified under the 
biaxial strain. A comprehensive study of the effect of strain would be promising, 
which can provides detail information on the responses of germanene under biaxial strain and explores the possible
typical properties. Hence, in this paper, based on first-principles calculations, we investigate the modification 
on the electronic structures and stability via phonon spectrum under the biaxial tensile strain for germanene. 
The phonon spectrum shows that germanene lattice can be stable up to 16\% biaxial tensile strain. We also calculate 
the Gr\"uneisen parameter and we find that the trend remain similar to silicene and behaves differently to graphene. 
The obtained results conclude that the biaxial tensile strain could bring an interesting $p$-doping phenomenon in germanene, 
which is consequences of the buckled structures and can not be possible in graphene.

We carried out first-principles calculations using density functional theory as implemented in the  QUANTUM-ESPRESSO 
package \cite{paolo}. A full relativistic Rappe-Rabe-Kaxiras-Joannopoulos type \cite{rrkjus} norm-conserving 
pseudopotential is employed together with the generalized gradient approximation in the Perdew, Burke, and 
Ernzerhof parametrization in order to include the spin orbit coupling. A Grimme scheme with scaling parameter 
0.75 is considered in the calculations to include the van der Waals interaction \cite{grime,kaloni-jmc}. The calculations 
are performed with a plane wave cutoﬀ energy of 60 Ryd. A Monkhorst-Pack 16$\times$16$\times$1 k-mesh is employed 
for optimizing the crystal structure and calculating the electronic band structure. Moreover, we use 
a 24$\times$24$\times$1 k-mesh to calculate the phonon spectrum. The atomic positions are relaxed until
an energy convergence of 10$^{-9}$ eV and a force convergence of 10$^{-4}$ eV/\AA\ are achieved. 
To avoid artifacts of the periodic boundary conditions we use an interlayer spacing of 15 \AA. The magnitude of the biaxial 
tensile strain is expressed as $\varepsilon = \frac{(a-a_0)}{a_0} \times 100\%$, where $a$ and $a_0 = 4.06$ \AA\ are the lattice 
parameters for strained and unstrained germanene, respectively.


Strain is the efficient way to engineered the electronic properties of graphene \cite{Pereira}. This allows 
to generate an all-graphene circuit, where all the elements of the circuit are made of graphene with 
different amounts and kinds of strain. It is also reported that the amounts and kinds of strain are 
equivalent to the magnetic field \cite{guinea}, which indeed, produce pseudo-magnetic quantum Hall effect. 
Other way around, for graphene up to 10\% strain is easily achievable \cite{Andresa}. It enhances the 
reactivity of graphene about 5 times as a result H atoms are bound strongly to the strained graphene. 
Which is the important route for H storage in graphene. We expect similar effect could be applicable for germanene because of 
its 2D structure similar to graphene. Therefore, we study in the following the effects of strain on the electronic structure 
and phonon spectrum. We obtain a lattice parameter of $a_0 = 4.06$ \AA\ and a buckling of 0.68 \AA\ for unstrained germanene. 
These structural parameters are in good agreement with previous reports \cite{Ni,ciraci}. We address the dependence of the 
stress (in Gpa) on the applied strain (in \%). The result is shown in Fig.\ 1. We find that the stress increases 
monotonically with the strain of 16\% and remain constant up to 19\% and decreases thereafter. Which in fact indicates 
that germanene is stable up to 16\% biaxial tensile strain, which is very similar to silicene with a similar 
buckled geometry. 
\begin{figure*}[h]
\includegraphics[width=0.5\textwidth,clip]{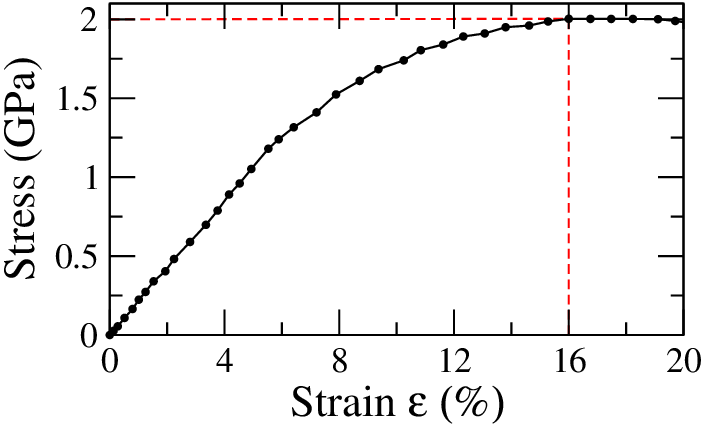}
\caption{Variation of the stress as a function of the applied biaxial tensile strain.}
\end{figure*}

In this section, we focus on the electronic band structure of germanene without/with variable biaxial tensile strain. 
We addressed the electronic band structure of the free standing germanene in Fig.\ 2(a). We have included spin orbit coupling 
in our calculations. It is noted that germanene behaves like a semiconductor with a band gap of 24 meV at the K point, see inset 
of Fig.\ 2(a), consistent with the previous findings \cite{Liu,Liu1}. The $\pi$ and $\pi^*$ bands of the Dirac cone are contributed 
by the $p_z$ orbitals of Ge, like as graphene and silicene. By the application of an external electric field, the magnitude of
the band gap could be enhanced easily because this field breaks the sublattice symmetry as a result the band gap due to spin orbit 
coupling could be increases. Such a effect has been observed in case of silicene \cite{Ni}. Due to its flexibility in the band gap 
opening, it can have a potential candidate in nano-electronic devices applications. 

Note that the the band gap of 24 meV in unstrained germanene becomes smaller (23 meV) for increasing strain. 
The reason is that the strain weakens the internal electric field because it reduces the magnitude of the buckling, 
which in fact reduces the strength of the intrinsic spin orbit coupling and thus the induced band gap is reduced. We find the Ge$-$Ge 
bond length is growing  monotonically with the strain as a result the buckling decreases. For unstrained germanene 
the Ge$-$Ge bond length of 2.44 \AA\ and buckling of 0.68 \AA\ are obtained. For 10\% strain these values change to 
2.65 \AA\ and 0.59 \AA, and for 16\% strain to 2.76 \AA\ and 0.55 \AA. The data for the variation of the Ge$-$Ge bond 
length and buckling under the biaxial strain are addressed in Table I.

\begin{table*}[ht]
\begin{tabular}{|c|c|c|c|c|c|c|c|c|}
\hline
$\varepsilon$ (\%) &Ge$-$Ge& $\Delta(\AA)$&$\theta^{\circ}$ &$\Delta\omega_G$ (cm$^{-1}$) &${\gamma_G}$&$s$ & $p$\\
\hline
5                  & 2.55 &0.63         & 112             &363.2                        &1.50 &1.55 &2.44\\
\hline
10                 & 2.65 &0.59         & 114             &303.6                        &1.45 &1.62 &2.36\\
\hline
16                 & 2.76 &0.55         & 115             &243.8                        &1.43 &1.69 &2.29\\
\hline  
20                 & 2.85 &0.51         & 116             &197.8                        &1.34 &1.75 &2.22\\
\hline
\end{tabular}
\caption{Strain, bond length, buckling, angle, and occupations.}
\end{table*}

We define that the $p$-doped Dirac states by the shift of the Dirac cone with respect to the Fermi level under biaxial tensile strain. 
The calculated band structure addressed in Fig.\ 2(a-b) shows that the Dirac cone shifts towards the higher energy 
range with respect to the the Fermi level by inducing $p$-doped Dirac states. At a strain of 5\%, we obtain a 
shifts of Dirac cone by 0.24 eV towards the higher energy range with respect to the Fermi level with a 23 meV band gap due to
intrinsic spin orbit coupling, see inset of Fig.\ 2(b). The intrinsic spin orbit gap is decreases by 1 meV due to 
the decreasing the buckling and hence electric field become weaker, such effect already had been found for strained 
silicene \cite{kaloni}. The conduction band at the $\Gamma$-point shifts towards the low energy range with respect to 
the Fermi level by 0.6 eV. This observation is well agree with the strained silicene \cite{Liu2} and germanene \cite{wang}. 
Note that due to shifts the $\Gamma$-point towards the higher energy range with respect to the Fermi level by leaving $p$-doped 
Dirac sates, consistent with the recent observation for silicene \cite{kaloni} and germanene \cite{wang}, which is 
in contrast to graphene. This can be attributed due to the fact that graphene is planar structure as compared to 
silicene and germanene and thus changes the $s-p$ hybridization significantly in later case. The occupation of $s$ and $p$ orbitals 
changes for unstrained and strained germanene, which in fact reduce the hybridization of the $s$ and $p$ orbitals. 
For unstained germanene the occupation at $s$ and $p$ orbitals are 1.47 e and 2.51 e, while for a strain of 5\% the 
occupation of $s/p$ orbitals increases/decreases (1.54/2.43 e), see Table I. The amount of the $p$-doped Dirac states 
are enhanced for increasing strain. The conduction band minimum at the $\Gamma$-point 
shifts further downwards and becomes more and more flat and occupied by increasing density of states at the Fermi level. 
At the strain of 10\%, the Ge$-$Ge bond length is increases, buckling is decreases, and hence angle is increases. This is a consequence 
of the weakening of Ge$-$Ge bonds strength. 

\begin{figure*}[h]
\includegraphics[width=0.5\textwidth]{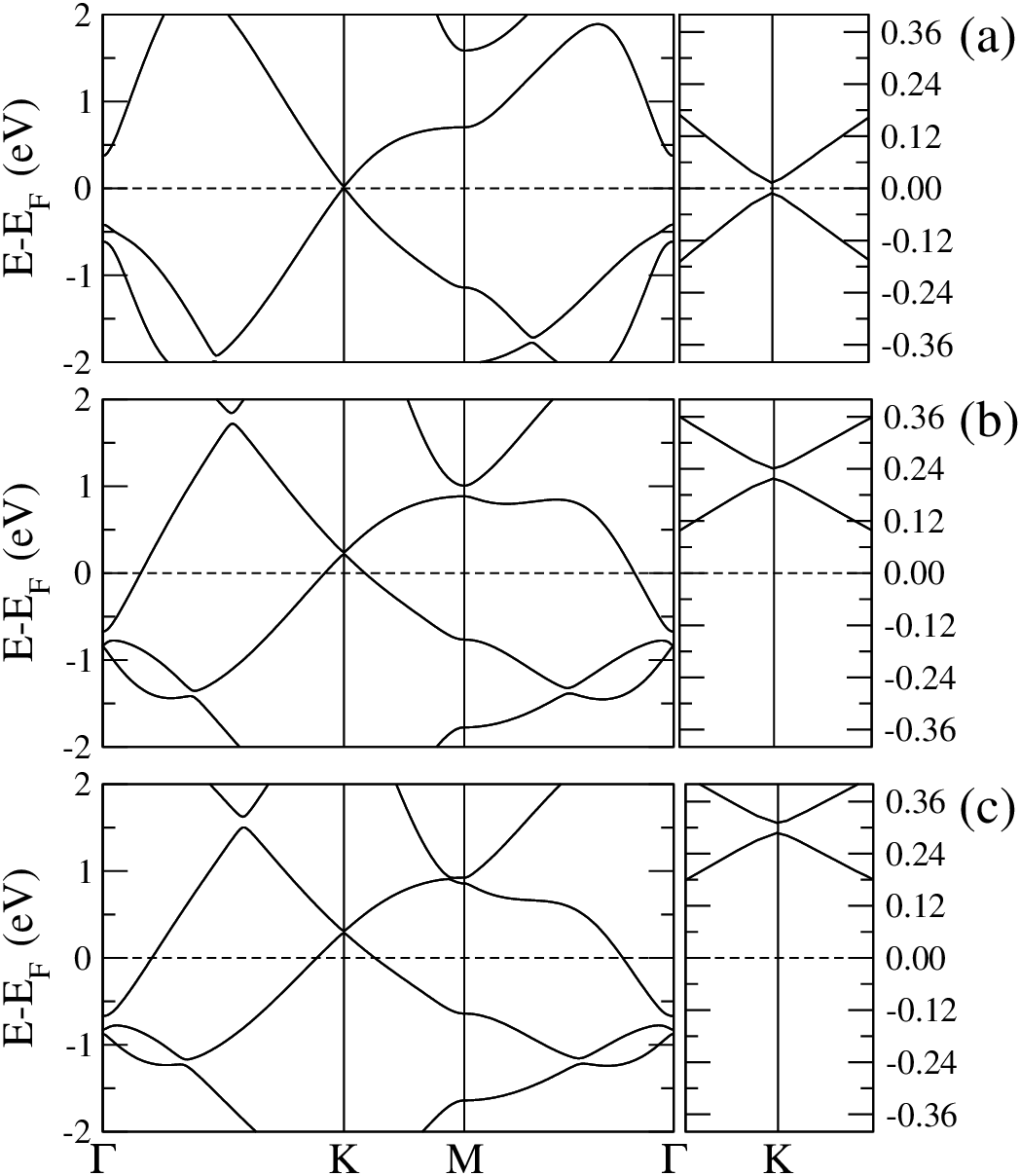}
\caption{Electronic band structure and partial densities of states for (a) unstrained germanene and germanene under biaxial tensile strain of (b) 5\% and (c) 16\%.}
\end{figure*}

The Dirac point lies at 0.3 eV towards higher energy range with respect to the Fermi level for strain of 16\%, see Fig. 2(c). The 
$\pi$ and $\pi^*$ bands of the Dirac cone are due to the $p_z$ orbitals of the Ge atoms with a minute contribution 
from the $p_x$ and $p_y$ orbitals, as it is expected. We obtain a gap of 22 meV, which in fact reduces as compared to unstrained 
germanene. The main reason for reducing the gap is the reduction of the buckling significantly (0.55 \AA) with increasing 
bond length of 2.76 \AA, and bond angle of $115^{\circ}$. The reduction of the buckling weakening the in built electric field 
as a result band gap is reduced, consistent with the strained silicene \cite{kaloni}. For higher strain the conduction band 
minimum shifts further towards lower energy range and the Dirac cone accordingly to higher energy range with respect to Fermi level. 
This is due to the change in the occupation in the $s$ and $p$ orbitals, see Table I. Since, the number of the electrons in the
system remain constant as a result the bands at the K/K$'$-points are depopulated and at the $\Gamma$-points are populated. 
Such a behavior is well agree with silicene but different from graphene, because the Ge$-$Ge bonds are much more flexible
than the C$-$C bonds in graphene. Contrary to silicene and germanene, the electronic structure of graphene does not changes 
in the presence of strain, resulting a zero band gap semiconductor up to a very large strain (30\%) \cite{choi}. 
This indicates that there is not any possibilities to achieve $p$-doping in graphene by strain. However, it has been 
demonstrated that a $p$-doping can be achieved in graphene by the intercalation of F and Ge with the SiC substrate \cite{kaloni-epl,cheng-apl}. 
The lattice becomes instable beyond strain of 16\% (for 20\% strain the parameters are presented in Table I), we will 
prove this fact by performing phonon calculation in the next section.

\begin{figure*}[h]
\includegraphics[width=0.5\textwidth,clip]{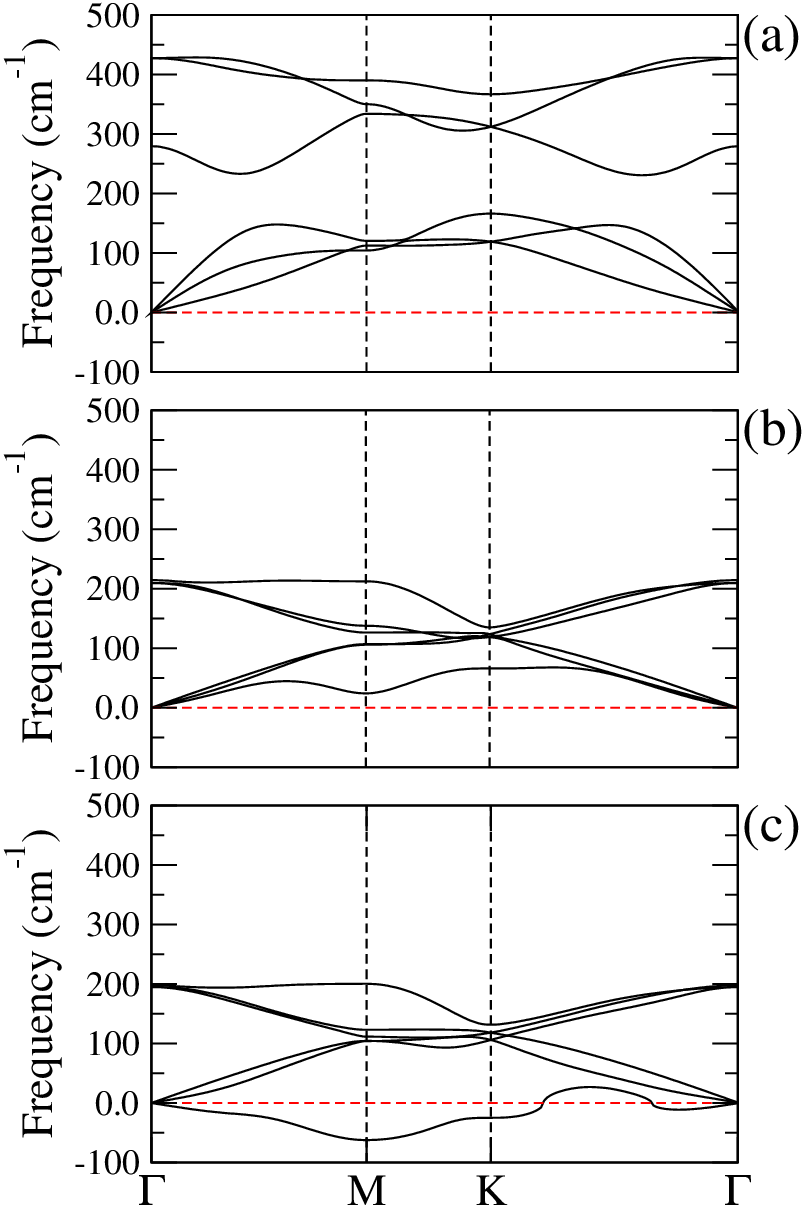}
\caption{Phonon frequencies for (a) unstrained germanene and germanene under biaxial tensile strain of (b) 16\% and (c) 20\%.}
\end{figure*}

In this section, we discuss the phonon spectrum of germanene unstrained and under strain of 5\%, 10\%, 16\%, and 20\%, see Fig.\ 3. 
For unstrained germanene the obtained optical phonon frequencies are 3.7 times smaller than graphene (1580 cm$^{-1}$ \cite{zabel}) 
and 1.28 times smaller than silicene (550 cm$^{-1}$ \cite{kaloni}). This can be realized by the smaller force constant and
weaker Ge$-$Ge bonds as compared to C$-$C and Si$-$Si bonds. 
Graphene shows a common features in the Raman spectra called G and D peaks, around 1580 cm$^{-1}$ and 1360 cm$^{-1}$ \cite{zabel}. 
The G peak corresponds to the E$_{2g}$ phonon at the $\Gamma$-point of the Brillouin zone. The D peak correspond to the K-point 
Brillouin zone. On the based on our knowledge the the Raman spectra of germanene is unknown. So, we do believe that our study would be 
a reference for the experimental observation of the Raman spectra to get insight of Raman frequencies and identification of G 
and D peaks. We therefore focus on the G peak and D peak identification 
in our study. For unstrained germanene the calculated phonon frequencies of G and D peaks are 427 cm$^{-1}$ and 366 cm$^{-1}$, 
respectively, lower than that of silicene. For strained silicene a significant modification of the phonon
frequencies is observed. At 5\% strain the G and D peak frequencies found to be 363 cm$^{-1}$ and 287 cm$^{-1}$, respectively. 
Which reflects that the weakening of the Ge$-$Ge bond under biaxial tensile strain. It also can be understood by the fact that 
the optical bands shows a clear trend of softening, which is expected because Ge$-$Ge bond length increase uniformly. For the strain 
of 10\% (16\%) obtained phonon frequencies of G and D peak are 303 cm$^{-1}$ (246 cm$^{-1}$) and 212 cm$^{-1}$ (150 cm$^{-1}$), 
respectively. We conclude that the germanene lattice is stable up to strain of 16\% because we still have positive frequencies 
along the $\Gamma$-M-K-$\Gamma$ path of the Brillouin zone. The germanene lattice becomes instable for the strain beyond 16\%. 
For this purpose we calculate phonon spectrum for the strain of 20\% and find a frequency of $-$62 cm$^{-1}$, see Fig.\ 3(c). 
Which indicates that the lattice is instable.

The effect of strain in 2D systems can be efficiently studied by Raman spectroscopy \cite{zabel}. Since the strain modifies the crystal 
phonon frequencies. The rate of phonon mode softening or hardening described by the Gr\"uneisen parameter, which in 
fact determines the thermomechanical properties fo the system. The Gr\"uneisen parameter for G peak under biaxial 
strain is given by 
\begin{equation}
\gamma_G=-\Delta\omega_G/2\omega_G^0\varepsilon,
\end{equation}
where $\Delta\omega_G$ is the difference in the frequency for unstrained and strained germanene and $\omega_G^0$ is 
the frequency of the G peak in unstrained germanene. The Gr\"uneisen parameter is difficult to study under uniaxial 
strain due to the fact that it require the Poisson ratio, which in fact depends on the choice of the substrate \cite{Mohiuddi}. 
It is also reported that it is difficult to calculate the D and 2D  Gr\"uneisen parameters because under uniaxial strain 
the position of the Dirac cones changes. The biaxial tensile strain is suitable to calculation the Gr\"uneisen parameter 
because it does not depend on the Poisson ratio as well as the position of the Dirac cone does not changes \cite{Mohiuddi}. 
Experimentally, the Gr\"uneisen parameter of graphene under biaxial strain has been demonstrated \cite{ding}. 
Recently, in the Ref.\cite{kaloni}, the Gr\"uneisen parameter for silicene under biaxial tensile strain has been 
studied theoretically. Hence, we calculate the Gr\"uneisen parameter for germanene and compare with the graphene 
and silicene. We find that the calculated Gr\"uneisen parameter is decreasing with increasing the biaxial tensile 
strain. This variation is well agree with calculated Gr\"uneisen parameter for silicene with a strain of 5\% to 25\% 
\cite{kaloni}. For a biaxial tensile strain of 5\%, 10\%, and 20\%, the obtained Gr\"uneisen parameter are 1.50, 
1.43, and 1.34, respectively, for silicene those values are 1.64, 1.62, and 1.34, respectively. The slight lowering 
of the Gr\"uneisen parameter in germanene as compared to silicene can be attributed by the lowering of the respective 
phonon spectrum. However, for graphene this magnitude of the Gr\"uneisen paramete can be obtained by a very low 
strain of 0.2\% \cite{ding}. The another reason for the lowering of the Gr\"uneisen parameter with increasing strain 
is the buckling decreases monotonically with increasing the strain and Ge$-$Ge bond length. Such a behavior is 
essentially similar to silicene (buckled structure) and different from graphene (non-buckled structure). We call 
for an experimental observations for the confirmation our findings.

In summary, we have performed first-principles calculations using density functional theory to study the effect of 
biaxial tensile train in germanene lattice, electronic properties, and phonon frequencies, and Gr\"uneisen paramete. 
Our results show that up to 16\%  biaxial tensile strain germanene lattice is in stable and the Dirac cone 
shift towards the higher energy range with respect to Fermi level as a result $p$-doped Dirac states are achieved. 
The realization of the $p$-doped Dirac states is due to the weakening of the Ge$-$Ge bonds, well agree with strained 
silicene \cite{kaloni}. We further calculate the phonon spectrum to demonstrate that germanene is stable up to 16\% 
under biaxial tensile strain. The calculated Gr\"uneisen parameter found to be similar to silicene and different from 
graphene as latter is non-buckled structure. The positive phonon frequencies up to a tensile strain of 16\% 
indicates that the germanene lattice is stabile in this regime, while the lattice becomes highly instable for the strain 
beyond 16\%, due to negative frequencies come in to the picture.


\begin{thebibliography}{50}
\bibitem{geim} A. H. Castro Neto, F. Guinea, N. M. R. Peres, K. S. Novoselov, and A. K. Geim, Rev. Mod. Phys. {\bf 81}, 109 (2009).



\bibitem{Topsakal}S. Cahangirov, M. Topsakal, E. Akturk, H. Sahin, and S. Ciraci, Phys. Rev. Lett. {\bf 102}, 236804 (2009).

\bibitem{padova}P. De Padova, C. Quaresima, C. Ottaviani, P. M. Sheverdyaeva, P. Moras, C. Carbone, D. Topwal, B. Olivieri, A. Kara, H. Oughaddou, 
B. Aufray, and G. Le Lay, Appl. Phys. Lett. {\bf 96}, 261905 (2010).

\bibitem{vogt}P. Vogt, P. De, C. Quaresima, J. Avila, E. Frantzeskakis, M. C. Asensio, A. Resta, B. Ealet, and G. Le Lay,  Phys. Rev. Lett. {\bf 108}, 155501 (2012).

\bibitem{ozaki}A. Fleurence, R. Friedlein, T. Ozaki, H. Kawai, Y. Wang, and Y. Yamada-Takamura, Phys. Rev. Lett. {\bf 108}, 245501 (2012). 

\bibitem{falko}N. D. Drummond, V. Z\'olyomi, and V. I. Fal$'$ko, Phys. Rev B {\bf 85}, 075423 (2012). 

\bibitem{Ni}Z. Ni, Q. Liu, K. Tang, J. Zheng, J. Zhou, R. Qin, Z. Gao, D. Yu, and J. Lu, Nano Lett. {\bf 12}, 113 (2012).

\bibitem{Houssa} M. Houssa, G. Pourtois, V. V. Afanas\'ev, and A. Stesmans, Appl. Phys. Lett. {\bf 96}, 082111 (2010).


\bibitem{Liu1} C. C. Liu, H. Jiang, and Y. G. Yao, Phys. Rev. B {\bf 84}, 195430 (2011). 

\bibitem{Liu} C. C. Liu, W. X. Feng, and Y. G. Yao, Phys. Rev. Lett. {\bf 107}, 076802 (2011). 

\bibitem{ma}Y. Ma , Y. Dai, C. Niu, and B. Huang, J. Mater. Chem. {\bf 22}, 12587 (2012).



\bibitem{Mohiuddi}T. M. G. Mohiuddin, A. Lombardo, R. R. Nair, A. Bonetti, G. Savini, R. Jalil, N. Bonini, D. M. Basko, C. Galiotis, 
N. Marzari, K. S. Novoselov, A. K. Geim, and A. C. Ferrari, Phys. Rev. B {\bf 79}, 205433 (2009).

\bibitem{choi}S. M. Choi, S. H. Jhi, and Y. W. Son, Phys. Rev. B {\bf 81}, 081407 (2010).

\bibitem{kaloni}T. P. Kaloni, Y. C. Cheng, and U. Schwingenschl\"ogl, J. Appl. Phys. {\bf 113}, 104305 (2013).

\bibitem{Liu2}G. Liu, M. S. Wu, C. Y. Ouyang, and B. Xu, EPL {\bf 99}, 17010 (2012).

\bibitem{paolo}P. Giannozzi, S. Baroni, N. Bonini, M. Calandra, R. Car, C. Cavazzoni, 
D. Ceresoli, G. L. Chiarotti, M. Cococcioni, I. Dabo, A. Dal Corso, S. de Gironcoli, S. Fabris, 
G. Fratesi, R. Gebauer, U. Gerstmann, C. Gougoussis, A. Kokalj, M. Lazzeri, L. Martin-Samos, N. Marzari,
F. Mauri, R. Mazzarello, S. Paolini, A. Pasquarello, L. Paulatto, C. Sbraccia, S. Scandolo, G. Sclauzero, 
A. P. Seitsonen, A. Smogunov, P. Umari, and R. M. Wentzcovitch, J. Phys.: Condens. Matter {\bf 21}, 395502 (2009).

\bibitem{rrkjus} A. M. Rappe, K. M. Rabe, E. Kaxiras, and J. D. Joannopoulos, Phys. Rev. B {\bf 41}, 1227 (1990).

\bibitem{grime}S. Grimme, J. Comput. Chem. {\bf 27}, 1787 (2006).

\bibitem{kaloni-jmc}T. P. Kaloni, Y. C. Cheng, and U. Schwingenschl\"ogl, J. Mater. Chem. {\bf 22}, 919 (2012). 

\bibitem{Pereira} V. M. Pereira and A. H. Castro Neto, Phys. Rev. Lett. {\bf 103}, 046801 (2009). 

\bibitem{guinea}F. Guinea, M. I. Katsnelson, and A. K. Geim, Nat. Phys. {\bf 6}, 30 (2010). 

\bibitem{Andresa} P. L. de Andresa and J. A. Verg\'es, Appl. Phys. Lett. {\bf 93}, 123531 (2008).

\bibitem{ciraci}H. Sahin, S. Cahangirov, M. Topsakal, E. Bekaroglu, E. Akturk, R. T. Senger, and S. Ciraci, Phys. Rev. B {\bf 80}, 155453 (2009).

\bibitem{wang} Y. Wang and Y. Ding, Solid State Cummun.{\bf 155}, 6 (2013).

\bibitem{kaloni-epl}T. P. Kaloni, M. Upadhyay Kahaly, Y. C. Cheng, and U. Schwingenschl\"ogl,  EPL {\bf 99}, 57002 (2012).

\bibitem{cheng-apl}Y. C. Cheng, T. P. Kaloni, G. S. Huang, and U. Schwingenschl\"ogl, Appl. Phys. Lett. {\bf 99}, 053117 (2011).

\bibitem{zabel} J. Zabel, R. R. Nair, A. Ott, T. Georgiou, A. K. Geim, K. S. Novoselov, and C. Casiraghi, Nano Lett. {\bf 12}, 617 (2012).

\bibitem{ding} F. Ding, H. Ji, Y. Chen, A. Herklotz, K. Dorr, Y. Mei, A. Rastelli, O. G. Schmidt, Nano Lett. {\bf 10}, 3453 (2010).





\end{thebibliography}
\end{document}